\documentclass[aps,twocolumn,showpacs,superscriptaddress]{revtex4}

\usepackage{amsmath}
\usepackage{amssymb}
\usepackage{graphicx}
\usepackage{bm}

\newcommand{\eps} {\varepsilon}

\begin{document}
\title{Structural changes in block copolymers:
coupling of electric field and mobile ions}

\author{Yoav Tsori}
\email{yoav.tsori@espci.fr}
\affiliation{Laboratoire Mati\`ere Molle \& Chimie\\
Unit\`e Mixte de Recherche ESPCI - CNRS - ATOFINA - (UMR 167) \\
ESPCI - 10 rue Vauquelin,
 75231 Paris CEDEX 05, France}

\author{Fran\c{c}ois Tournilhac}
\affiliation{Laboratoire Mati\`ere Molle \& Chimie\\
Unit\`e Mixte de Recherche ESPCI - CNRS - ATOFINA - (UMR 167) \\
ESPCI - 10 rue Vauquelin,
 75231 Paris CEDEX 05, France}

\author{David Andelman}
\email{andelman@post.tau.ac.il}
\affiliation{School of Physics and Astronomy, Raymond and Beverly Sackler
Faculty of Exact Sciences\\
Tel Aviv University, 69978 Ramat Aviv, Israel}

\author{Ludwik Leibler}
\affiliation{Laboratoire Mati\`ere Molle \& Chimie\\
Unit\`e Mixte de Recherche ESPCI - CNRS - ATOFINA - (UMR 167) \\
ESPCI - 10 rue Vauquelin,
 75231 Paris CEDEX 05, France}

\date{26/7/2002}

\begin{abstract}

We consider the coupling between an external electric field and
dissociated ions embedded in anisotropic polarizable media such
as block copolymers. We argue that the presence of such ions can
induce strong morphological changes and even lead to a structural
phase transition. We investigate, in particular, diblock
copolymers in the body centered cubic (bcc) phase. In pure
dielectric materials (no free charges), a dielectric breakdown is
expected to occur for large enough electric fields, preempting
any structural phase transition. On the other hand,  dissociated
ions can induce a phase transition for fields of about $10$
V/$\mu$m or even lower. This transition is from an insulator
state, where charges are preferentially localized inside bcc
spherical domains, to a conducting state where charges can move
along the long axis of oriented cylinders (forming a hexagonal
phase). The cylinders diameter is small and is determined by the
volume fraction of the minority copolymer component. The strength
of this mechanism can be tuned by controlling the amount of free
ions present as is observed by our experiments. These theoretical
predictions  support  recent experimental findings on the bcc to
hexagonal phase transition in copolymer systems.
\end{abstract}


\maketitle

Designed control of material properties in the sub-micrometer
range has drawn considerable interest in recent years because of
its importance in applications as well as in basic research. As
possible applications we mention waveguides, photonic band gap
materials \cite{VBSN01} and dielectric mirrors \cite{FT98}. Recent
studies have highlighted the role of external electric fields in
creating well aligned structures. An electric field was used to
create an instability in  polymer films, thus replicating the
pattern of a master template \cite{1}. In block copolymers
(composed of several chemically distinct polymer sequences
connected together by covalent bonds), electric fields are
effective in aligning micro-domains in a desired direction, as has
been shown experimentally \cite{AH94,AH93} and theoretically
\cite{AH93,PW99,AM01,TA02}.  For example, removal of one of the
two   components of a copolymer film is used to produce
anti-reflection coatings for optical surfaces \cite{WS99}.
Furthermore, diblock copolymers, such as polystyrene
(PS)/polymethylmethacrylate (PMMA) in its hexagonal phase was used
as a starting point to produce an array of long, aligned and
conducting nano-wires \cite{RS00}.

A well known method to cause orientation or structural changes in
heterogeneous dielectric media is based on the ``dielectric
mechanism''. As an example, we consider a  material composed of
two types of micro-domains  having dielectric constants, $\eps_A$
and $\eps_B$, respectively, and placed in an external electric
field ${\bf E}$. An electrostatic energy penalty will be paid each
time that a dielectric interface is oriented perpendicular to the
field \cite{TA02,Onuki95,TDR00}. Thus, a state where $ {\bm \nabla}\eps$
is perpendicular to ${\bf E}$ is favored. This energy cost is
proportional to $(\eps_A-\eps_B)^2E^2$, and is enhanced when the
difference in polarizabilities  is large. A real concern is
whether this relatively weak effect (2nd order in $E$) requires
high fields exceeding the typical values for dielectric breakdown
of the entire film. This breakdown occurs for fields $\gtrsim 100$
V/$\mu$m in thick films of $1$-$3$ mm, but is expected to be much
smaller for thin films as considered here \cite{UPSH}.

So far, little  attention was given to the existence of free
(mobile) ions inside  dielectric copolymer materials. These ions
are found in large amount because during anionic polymerization,
the reaction is initiated by an organometallic reagent (usually
butyl lithium). When monomers are eventually exhausted, each
carbonionic chain  neutralized with water gives rise to one
neutral polymer chain and one metal hydroxide unit (usually LiOH).
The use of organolithic reagents and catalysts
 increases even further the total amount of ions \cite{Navarro98}.

Based on the importance of free ion coupling to external $E$
fields, we propose an alternative mechanism to explain
morphological changes and even phase transitions in block
copolymers. Similar considerations can apply in general to any
highly polarizable heterogeneous media. The new driving force is
the free energy gain of  free charges (dissociated ions) as they
move upstream or downstream the electrostatic potential (depending
on their sign). The fields needed to drive morphological changes
are typically much smaller than the fields resulting from  the
dielectric mechanism and their value is well below dielectric
breakdown threshold. Moreover, tuning the free ion concentration
offers a convenient  control of these morphological changes.

Although our approach is general, we concentrate here on  PS/PMMA
diblock copolymers used by several experimental groups
\cite{RS00}. We show that with the alternative ``free ions
mechanism" there is a phase transition from a body centered cubic
(bcc) lattice of spherical domains to a hexagonal array of
cylinders of small diameter, in agreement with recent experimental
findings of Russell and co-workers \cite{Rprivate02}. Because of
the ionic coupling with an external $E$ field, the cylinder
alignment is long range as is important in many of the
aforementioned applications.

In an A/B diblock copolymer melt the {\it macroscopic} phase
separation occurring for two A/B immiscible chains (or blocks) is
hindered because of chain connectivity. Instead, the system
undergoes a {\it mesoscopic} phase separation, with typical length
scales of dozens of nanometers. In the bulk, phase behavior is by
now fairly well understood \cite{Leibler80,OK86} and was found to
consist of several spatially modulated phases of different
symmetry. The Flory parameter $\chi\sim T^{-1}$ characterizes the
repulsion between the blocks and $f=N_A/N$ is the fraction of A
monomers in a copolymer chain of $N=N_A+N_B$ monomers. The phase
behavior (in mean-field theory) is given by $f$ and the combined
 $N\chi$ parameter. For small values of $N\chi$ (high temperature)
the system is in a mixed, disordered, phase. Lowering the
temperature or equivalently raising $N\chi$ above $N\chi_c\simeq
10.5$ results in a weak first-order phase-transition to a lamellar
phase for symmetric melts ($f=\frac12$). Increase of
$|f-\frac12|$ changes the spontaneous curvature and induces a
transition to a hexagonal phase of cylinders, a gyroid phase of
cubic symmetry or a bcc lattice of spherical domains.

The copolymer bulk free energy $F_b$ an be written as a functional
of the local order parameter $\phi({\bf r})=\phi_A({\bf r})-f$.
This order parameter denotes the deviation of the A monomer volume
fraction from its average value $\langle\phi_A\rangle=f$, and the
free energy expression is:
\begin{eqnarray}\label{Fbulk}
\frac{Nb^3F_b[\phi]}{k_BT}=
\nonumber\\
&&\hspace{-2cm}\int {\rm
d}^3r\left\{\frac12\tau\phi^2+\frac12h\left(q_0^2\phi+{\bf \nabla}^2
\phi\right)^2 +\frac16\Lambda\phi^3+\frac{u}{24}\phi^4\right\}
\end{eqnarray}
This  and  similar (mean-field) free energy forms have been used
extensively in the past to describe spatially modulated phases
\cite{Swift77,Andelman95} such as  weakly segregated diblock
copolymers \cite{TA02,Leibler80,OK86,FH87,TA01},  Langmuir films
\cite{ABJ84} and magnetic (garnet) films \cite{GD82}. In
Eq.~(\ref{Fbulk}), $\tau\sim N(\chi_c-\chi)$ is the reduced
temperature and $q_0\sim R_g^{-1}$ is the first wavenumber which
becomes stable upon transition to the lamellar phase. The lattice
periodicity $d_0=2\pi/q_0$ is about $10$ nanometers, $b$ is Kuhn
segment length,  $h\sim R_g^4$ and the constants $\Lambda$ and
$u$ are the 3-point and 4-point vertex functions calculated by
Leibler \cite{Leibler80} in the weak segregation limit. They
depend on $f$ and $\tau$.

Contrary to lamellar or hexagonal phases where orientation by
electric fields is related to a preferred spatial orientation
\cite{AH94,AH93,PW99,AM01,TA02,WS99,RS00,Onuki95,TDR00}, a bcc
lattice of spheres is isotropic and no orientation is possible. On
the other hand, strong enough electric fields will elongate the
spheres and eventually will cause a morphological changes of the
bcc phase into a hexagonal array of cylinders.

Let us consider first a PS-PMMA copolymer melt in the absence of
external $E$ field. The parameters $N\chi$ and $f$ are chosen
such that the thermodynamically stable phase is a bcc stack of
PMMA spheres embedded in a PS matrix. Within the single $q$-mode
approximation employed throughout this paper (weak segregation
limit), the order parameter $\phi$ for the bcc phase is given by
\begin{eqnarray}\label{phi-bcc}
\phi^{\rm bcc}({\bf r})=A_{\rm bcc}\sum_{n=1}^{6}\cos({\bf
q}_n\cdot{\bf r})
\end{eqnarray}
$A_{\rm bcc}$ is the amplitude of density modulations, and the
six fundamental $q$-modes are given by
\begin{eqnarray}
{\bf q}_{1,4}&=&q_0 (\mp 1,0,1)/\sqrt{2},~~~ {\bf
q}_{2,5}=q_0 (1,\mp 1,0)/\sqrt{2}\nonumber \\
{\bf q}_{3,6}&=&q_0 (0,1,\mp 1)/\sqrt{2}
\end{eqnarray}

Consider now only the dielectric mechanism. We assume for
simplicity that the external electric field E is applied along the
$(1,1,1)$ direction of the lattice: ${\bf E}=E(1,1,1)$. This
choice of  ${\bf E}$ direction  does not affect the magnitude of
the critical field $E_c$  \cite{remark1}. As the field is turned
on, the PMMA spheres elongate along the field direction, thereby
reducing the component of ${\bf \nabla}\eps$ along ${\bf E}$. We note
that the bcc phase can be regarded as a stack of $(1,1,1)$ planes,
each with hexagonal symmetry. For sufficiently strong fields, the
electrostatic energy dominates and the system is composed of
cylinders oriented along the $(1,1,1)$ direction with dielectric
interfaces parallel to $E$. The order parameter can now be
written as a sum of two terms. The first contains the $q_1$, $q_2$
and $q_3$ modes and has a {\it hexagonal} symmetry, whereas the
second contains the $q_4$, $q_5$ and $q_6$ modes. Hence, as a
function of electric field strength the order parameter $\phi({\bf
r},{\bf E})$ is:
\begin{eqnarray}\label{ansatz}
\phi({\bf r},{\bf E})=w(E)\sum_{n=1}^{3}\cos({\bf q}_n\cdot{\bf
r})+g(E)\sum_{m=4}^{6}\cos({\bf q}_m\cdot{\bf r})
\end{eqnarray}
For large enough field the order parameter reduces to the
hexagonal one
\begin{equation}\label{phi-hex}
\phi^{\rm hex}({\bf r})=A_{\rm hex}\sum_{n=1}^{3}\cos({\bf
q}_n\cdot{\bf r})
\end{equation}
where $A_{\rm hex}$ is the amplitude of density modulations in
the hexagonal phase. Clearly, $A_{\rm hex}$ is the limit of
$w(E)$ for large E-fields, while $g(E)$ should vanish there.

In the weak segregation regime one can expand the electrostatic
free energy per unit volume in small density variations $\phi$ and
obtain the following form \cite{AH94,AH93,TA02,Onuki95}:
\begin{eqnarray}
\frac{Nb^3}{k_BT}F_{\rm dielec}=\beta\sum_q (\hat{{\bf
q}}\cdot{\bf E})^2\phi_{\bf q}\phi_{-\bf q}=\beta
E^2g^2\label{Fel}
\end{eqnarray}
where $\beta=Nb^3(\eps_{_{\rm PS}}-\eps_{_{\rm
PMMA}})^2/(2\bar{\eps} k_BT)$ \cite{AH93}, the static dielectric
constants of PS and PMMA for high temperatures ($>160^\circ$ C)
 are $\eps_{_{\rm
PS}}\approx 2.5$ and $\eps_{_{\rm PMMA}}\approx 6$, respectively,
and $\bar{\eps}=f\eps_{_{\rm PMMA}}+(1-f)\eps_{_{\rm PS}}$.

The free energy $F_b+F_{\rm el}$ [Eqs. (\ref{Fbulk}) and
(\ref{Fel})] is then minimized with respect to $w(E)$ and $g(E)$
for given parameters $\tau$, $\Lambda$, $u$ and electric field
$E$. The results of this minimization are shown in Fig. 1. When
the field strength is $E=0$,  $w(0)=g(0)=A_{\rm bcc}$. As $E$
increases from zero, $w(E)$ increases monotonically while $g(E)$
decreases monotonically. At a {\it critical} field, $E=E_c$,
$w(E)$ jumps discontinuously to $A_{\rm hex}$, while $g(E)$ jumps
(down) to zero. For all fields $E>E_c$ the system is in a state
with hexagonal symmetry [$g(E)=0$].

%----------------------------------------------
\begin{figure}[h!]
\begin{center}
\includegraphics[scale=0.45,clip]{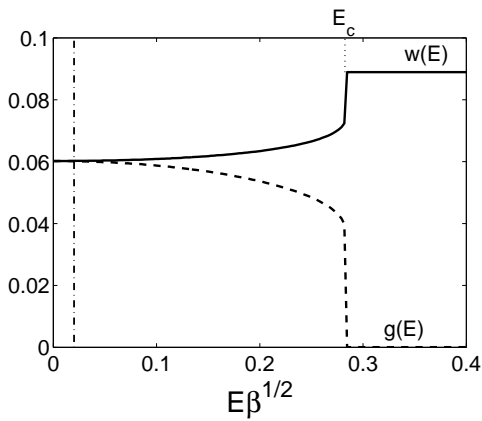}
\end{center}
\label{Fig. 1} \caption{Weight amplitudes $g(E)$ and $w(E)$ [Eq.
(\ref{ansatz})] as function of external E-field in dielectric
mechanism. The transition to bcc phase occurs at the critical
field $E_c\approx 70$ V/$\mu$m where these amplitudes are
discontinuous. Dash-dot line is the reduced transition field
$\approx 6$ V/$\mu$m of the free ions mechanism. The chosen
parameters are $f=0.37$ and $N\chi=12$, $N=500$, $b=2.5\cdot
10^{-10}$ m, $\beta=1.7\cdot 10^{-17}$ m$^2/$Volt$^2$. Explicit
dependence of $\tau$, $h$, $u$ and $\Lambda$ on $N\chi$ and $f$ is
given in Ref.~\cite{FH87}.}
%given in Ref.~[20].}
\end{figure}

%%%%%%%%%%%%%%%%%%%%%%%%%%%%%%%%%%%%%%%%%%%%%%%%%%%%%%%%%%%%%%%%%%%%%55

The inter material dividing surface (IMDS) is defined as the
surface where the A and B monomer densities are equal and is
given by $\phi({\bf r})=\frac12-f$. The IMDS is shown in Fig.
2~(a) for the bcc phase with $E=0$. In Fig. 2~(b) the field is
slightly below $E_c$, $E=0.98E_c$, and the spheres are
substantially elongated. In part (c) $E$ is slightly above the
transition, $E=1.02E_c$, and the system is composed of cylinders
oriented along the field direction. The value of $E_c$ for
PS/PMMA is estimated to be quite large, $E_c\approx 70$ V/$\mu$m.
It is comparable to the value for dielectric breakdown and much
bigger than the reported bcc to hexagonal transition fields
\cite{Rprivate02}.

%%%%%%%%%%%%%%%%%%%%%%%%%%%%%%%%%%%%%%%%%%%%%%%%%%%%%%%%%%%%%%%55
\begin{figure}[h!]
\begin{center}
\includegraphics[scale=0.5,bb=50 35 335 745,clip]{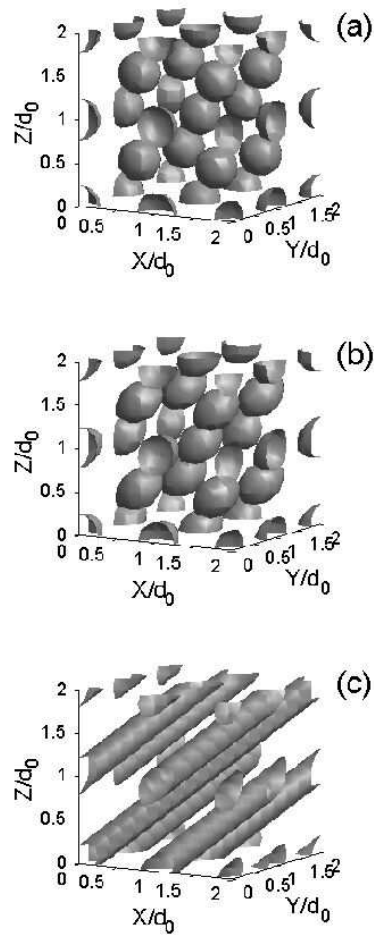}
\end{center}
\label{Fig. 2} \caption{Plot of the inter material dividing
surface (IMDS). A bcc lattice of spheres with $E=0$ is shown in
(a). In (b) the field is $E=0.98E_c$ and the spheres are highly
deformed.  The system undergoes an abrupt change into an hexagonal
phase of cylinders for $E=E_c$. In (c) the shown field is
$E=1.02E_c$. The parameters are the same as in Fig. 1.}
\end{figure}
%%%%%%%%%%%%%%%%%%%%%%%%%%%%%%%%%%%%%%%%%%%%%%%%%%%%%%%%%%%%%%%%%55

In order to resolve the discrepancy between experiments and the
dielectric mechanism,  we now turn to calculate the free charge
effect since ions are present in the BCP melt. In PMMA the ionic
solvating ability is due to Li$^+\leftarrow$O=C coordination
bridges. Nonpolar polymers such as PS do not form coordination
bridges of this type and preclude the dissociation of ionic pairs.
Hence, dissociated charges are found mainly inside the PMMA-rich
spherical domains, as is measured by Kim and Oh \cite{Kimoh00} and
reconfirmed by our own experiments \cite{tobepub}.
 We denote by $Q$ the total
charge of {\it dissociated} ions inside each PMMA spherical
domain. $Q$ is much smaller than the nominal amount of ions in the
sphere. It should be emphasized that in contrast to the ``leaky
dielectric'' model of Taylor and Melcher \cite{Taylor66}
describing the electrohydrodynamics phenomena in conducting
dielectric fluids, here there is no flow of material, but rather
an elastic behavior of the polymer chains.

We carried out DC conductivity measurements at $160 ^\circ$ C in
pure PMMA and PS films doped with $6.6\cdot 10^{14}$ LiOH ions per
m$^3$ \cite{tobepub}. In the PMMA sample, the integrated transient
current after the voltage is applied   indicates that the mobile
(dissociated) ions constitute a fraction of $3\cdot 10^{-5}$ of
the total embedded ions. The actual fraction of dissociated pairs
in the PS/PMMA block copolymer system is expected to be higher
because of the high doping \cite{Kimoh00}. The hazy and strongly
birefringent PS sample showed no decaying currents, indicating a
negligible amount of dissociated pairs. Since each spherical
domain has approximately $6$ PMMA chains, we deduced from the
fraction $3\cdot 10^{-5}$ mentioned above that $Q\gtrsim 10^{-3}
e$. In addition, from the conductivity data and charge density
estimates we calculate the Li$+$  mobility in PMMA to be
$\mu\simeq 4.6\cdot 10^5$ m$^2$/J~sec.

In presence of the $E$-field, there is $+Q$ charge at one end of
the bcc domain and $-Q$ at the other end,  creating an effective
dipole. The two energy contributions are the dipole-field and
dipole-dipole interactions. The first contribution is the usual
dipole-field interaction. The dipole moment of each bcc domain is
$|{\bf p}|=2QR$, where $R\approx 0.2d_0\approx 2$nm is the sphere
radius. The dipole-field interaction energy per unit volume in
the bcc phase is
\begin{equation}
F^{\rm bcc}_{\rm d-f}=-4QRE/d_0^3
\end{equation}
Note that this energy is linear in the field  $E$ and in the
charge $Q$.

In the hexagonal phase the drift distance $\lambda$ for
dissociated ions can be very large. For an AC field of frequency
$w$, $\lambda$ is given by $\lambda=2\pi e\mu E/w$. In a DC field
($w=0$) this infinite length is limited by the experimental
duration (dozens of seconds) or the system size, and therefore we
take $\lambda\approx 1$ $\mu$m (which corresponds to time scales
of about 1 sec). The dipole-field interaction in the hexagonal
phase is
\begin{equation}
F^{\rm hex}_{\rm d-f}=-4Q\lambda E/d_0^3\gg F^{\rm bcc}_{\rm d-f}
\end{equation}
The ability of the charges to delocalize and diffuse in the
hexagonal phase is the driving force behind this bcc to hexagonal
transition.

The dipole-dipole interaction can be divided to interaction of
the charges inside the same sphere and between neighboring
spheres. Inside the same domain the interactions between charges
are balanced by thermal dissociation and, therefore, only affect
their number $Q$. Dipole-dipole interaction between adjacent
domains is estimated by summing up all dipoles taken to be
parallel to the field. The contribution per unit volume is
\cite{SK98}
\begin{equation}
F_{\rm d-d}=\frac{10\pi}{3}\frac{(2QR)^2}{\eps_{\rm eff}d_0^6}
\end{equation}
where $\eps_{\rm eff}$ is the effective dielectric constant of
the compound material and is taken here to be just the average,
$\bar{\eps}$. This free energy contribution is positive
(destabilizing the bcc phase) and  quadratic in the charge $Q$.

We now compare the total free energy difference between the bcc
and hexagonal phases.
 The last terms calculated, $F_{\rm d-d}$ and
$F^{\rm hex}_{\rm d-f}-F^{\rm bcc}_{\rm d-f}\approx F^{\rm
hex}_{\rm d-f}$ need to be compared to the dielectric contribution
 $F_{\rm dielec}$ and  the bare
polymeric difference $\Delta F=F_b[\phi^{\rm hex}]-F_b[\phi^{\rm
bcc}]\approx {k_BT}/(200{Nb^3})$ [see Ref. \cite{FH87} or Eqs.
(\ref{Fbulk}), (\ref{phi-bcc}) and (\ref{phi-hex})]. The value of
$\Delta F$ can be reduced by approaching the bcc/hex phase
transition boundary on the $N\chi$--$f$ phase diagram. When
comparing orders of magnitudes for a field $E=6$ V/$\mu$m we see
that
\begin{equation}
\frac{F^{\rm hex}_{\rm d-f}}{\Delta F}\simeq 1~,~~~~ \frac{F_{\rm
dielec}}{\Delta F}\simeq 6\cdot 10^{-4}~,~~~~\frac{F_{\rm
d-d}}{\Delta F}\simeq 3\cdot 10^{-5}
\end{equation}
Therefore, the dominant mechanism in the bcc to hexagonal
transition is the interaction of ions with the external fields,
and for the above parameters it is about $10^4$ stronger than the
dielectric mechanism. Because $F_{\rm dielec} \sim E^2$  [Eq.
(\ref{Fel})] while $F_{\rm d-f}^{\rm hex}\sim E$, only at much
higher $E$-fields the former will dominate. The dipole-dipole
interaction between domains is independent on $E$ and is
negligible.

In conclusion, we propose an alternative mechanism explaining
structural changes in polarized media such as diblock copolymers.
The mechanism is based on the interaction of free charges with an
external field rather than on the tendency to reduce dielectric
interfaces perpendicular to the field. We calculate the
deformation of the bcc phase of diblock copolymers using the
dielectric mechanism only, and find that the critical field for
transition to a hexagonal phase is $E_c\approx 70$ V/$\mu$m. It
is shown that the ions which are normally present in these
materials can cause a similar structural change but with a much
smaller field of order $6$ V/$\mu$m. Thus, this mechanism should
be the dominant one. Moreover, the strength of this mechanism can
be greatly enhanced by addition of ions. We note that there has
been already some evidence that the bcc to hexagonal transition in
diblock copolymers can be achieved experimentally with such small
$E$-field \cite{Rprivate02}. It will be of interest to further
investigate the effect of these free ions on other morphological
changes and structural phase transitions in anisotropic
polarizable media.

\bigskip
{\it Acknowledgements.~~~~} We thank T. Russell for discussions
and communication of his experimental results prior to
publication. We thank E. Kramer and T. Thurn-Albrecht for useful
comments. Partial support from the U.S.-Israel Binational
Foundation (B.S.F.) under Grant 98-00429, the Israel Science
Foundation founded by the Israel Academy of Sciences and
Humanities Centers of Excellence Program, and the Chateaubriand
fellowship program is gratefully acknowledged.
%%%%%%%%%%%

%

%------------------------------------------------


\begin{thebibliography}{0}
\bibitem{remark1} A calculation with an arbitrary direction
of the field $E$ and a different order parameter ansatz in Eq.
(\ref{ansatz}) was carried out (not shown here) without
significantly changing the value of the  $E_c$ field for the
 for the bcc to a hexagonal transition.
\end{thebibliography}
\end{document}